\documentclass[aps,onecolumn,showpacs,superscriptaddress,nofootinbib,notitlepage]{revtex4-1} 
\usepackage{graphics,graphicx} 
\usepackage{dcolumn}   
\usepackage{bm}        
\usepackage{amssymb}   
\usepackage{bigints}
\usepackage{xcolor}
\usepackage{xspace}
\usepackage{tabularx}
\usepackage{hyperref}
\usepackage{multirow}

\def\be{\begin{equation}}
\def\ee{\end{equation}}
\def\bea{\begin{eqnarray}}
\def\eea{\end{eqnarray}}

\newcommand{\eV}{\ensuremath{{\mathrm{\,e\kern -0.1em V}}}\xspace}
\newcommand{\kev}{\ensuremath{{\mathrm{\,ke\kern -0.1em V}}}\xspace}
\newcommand{\mev}{\ensuremath{{\mathrm{\,Me\kern -0.1em V}}}\xspace}
\newcommand{\gev}{\ensuremath{{\mathrm{\,Ge\kern -0.1em V}}}\xspace}
\newcommand{\gevsq}{\ensuremath{{\mathrm{\,Ge\kern -0.1em V}^2}}\xspace}
\newcommand{\tev}{\ensuremath{{\mathrm{\,Te\kern -0.1em V}}}\xspace}
\newcommand{\nsgev}{\ensuremath{{\mathrm{Me\kern -0.1em V}}}\xspace}
\newcommand{\nsmev}{\ensuremath{{\mathrm{Me\kern -0.1em V}}}\xspace}
\newcommand{\nskev}{\ensuremath{{\mathrm{ke\kern -0.1em V}}}\xspace}

\usepackage{mfirstuc} 
\newcommand{\addReviewer}[2]{
  \expandafter\newcommand\csname #1\endcsname[1]{{\textbf{ \color{#2} \capitalisewords{#1}:\,##1}}}
  \expandafter\newcommand\csname #1cor\endcsname[2]{{\color{#2} \capitalisewords{#1}:\,\st{##1}{\textbf{##2}}}}
  \expandafter\newcommand\csname #1color\endcsname{#2}
  \expandafter\newcommand\csname #1todo\endcsname[1]{{\todo[inline,color=white!70!#2, caption={}]{\textbf{\capitalisewords{#1}}: ##1}}}
}

\usepackage{soul,color}
\definecolor{chromeyellow}{rgb}{1.0, 0.65, 0.0}
 \addReviewer{ale}{red}

\begin{document}
\title{Doubly Heavy Tetraquarks in the  Born-Oppenheimer approximation}

\author{Luciano Maiani}
\affiliation{Dipartimento di Fisica and INFN, Sapienza Universit\`a di Roma, Piazzale Aldo Moro 2, I-00185 Roma, Italy}
\affiliation{CERN, 1211 Geneva 23, Switzerland}

\author{Alessandro Pilloni}
\affiliation{Universit\`a degli Studi di Messina, Viale Ferdinando Stagno d'Alcontres 31, I-98166 Messina, Italy}
\affiliation{INFN Sezione di Catania, Via Santa Sofia 64, I-95123 Catania, Italy}

\author{Antonio D. Polosa}
\affiliation{Dipartimento di Fisica and INFN, Sapienza Universit\`a di Roma, Piazzale Aldo Moro 2, I-00185 Roma, Italy}

\author{Veronica Riquer}
\affiliation{Dipartimento di Fisica and INFN, Sapienza Universit\`a di Roma, Piazzale Aldo Moro 2, I-00185 Roma, Italy}
\affiliation{CERN, 1211 Geneva 23, Switzerland}

\date{\today}
\begin{abstract}
Tetraquarks $QQ\bar q\bar q$ are found to be described remarkably well with the Quantum Chromodynamics version of the Hydrogen bond, as treated with the
Born-Oppenheimer approximation. We show the robustness of the method by computing the mass of the observed ${\cal T}_{cc}$ tetraquark following two different paths. 
Relying on this, we provide a  prediction for the mass of  the expected ${\cal T}_{bb}$ particle. The average sizes of tetraquarks are estimated to be approximately $3$--$5\gev^{-1}$. As a consequence hyperfine separations are not expected to be sizeable. We discussed possible reasons why LHCb has observed only one state in the $DD^*$ spectrum.
\end{abstract}


\maketitle


{\bf \emph{Introduction.}}  The discovery of a doubly charm meson~\cite{LHCb:2021vvq,LHCb:2021auc}, as well as the theoretical consensus on the existence of a doubly bottom counterpart~\cite{Bicudo:2012qt,Karliner:2013dqa,Francis:2016hui,Eichten:2017ffp,Leskovec:2019ioa}, is moving the spotlight on 
heavy-light $QQ\bar q \bar q$ tetraquarks. Since they cannot mix with ordinary charmonia, they turn out to be the simplest exotic system to study, see~\cite{Esposito:2013fma}.

Given the separation of masses $M_Q\gg m_q$, one finds a situation similar to that encountered in the hydrogen molecule.  The fast motion of the light quarks in the field of the heavy color sources generates an effective potential, dependent on the relative distance $R$ separating the $QQ$ pair. The potential, in turn, regulates the slower motion of the heavy quarks.  Such an effective potential, known as the Born-Oppenheimer potential (BO),
is obtained by solving the eigenvalue equation for the light particles at fixed values of the coordinates of the heavy particles~(see e.g. \cite{Braaten:2014qka,Brambilla:2017uyf,Bicudo:2017szl,Giron:2019bcs,Prelovsek:2019ywc}). The energy ${\cal E}$ will be a function of the relative distance $R$ between heavy particles and corresponds to  the core of the full BO potential, which includes the direct interaction between the sources. 

When solving the Schr\"odinger equation of the heavy particles, one neglects the momentum of the heavy particles computed as the gradient of the eigenfunction related to ${\cal E}$. This is the content of the {\it Born-Oppenheimer approximation}, illustrated in detail for QED in~\cite{weinbergQM,pauling}.
 
Recently, we have applied the Born-Oppenheimer approximation to calculate the mass of the doubly charm baryon $\Xi_{cc}$ and of the lowest lying doubly  heavy tetraquarks, ${\cal T}_{cc}$ and ${\cal T}_{bb}$~\cite{Maiani:2019lpu}. In synthesis, the calculation gave a mass of $\Xi_{cc}$ in reasonable agreement with observation, but a mass of ${\cal T}_{cc}$ close to the $D D$ threshold and a mass for ${\cal T}_{bb}$ considerably below the $ B B$ threshold, deep in the stability region against weak and electromagnetic decays. 
Previous calculations based on constituent quark model~\cite{Karliner:2017qjm,Eichten:2017ffp,Luo:2017eub,Hernandez:2019eox,Guo:2021yws} 
had rather indicated a ${\cal T}_{cc}$ mass close to the $D D^*$ threshold and, for ${\cal T}_{bb}$, a $Q$-value well inside the stability region. 

The observation of ${\cal T}_{cc}(3875)^+$ at the $D D^*$ threshold calls for a closer examination of our calculation~\cite{Maiani:2019lpu}. 
As emerging from recent debates on the compositeness of exotic states~\cite{Esposito:2021vhu,Baru:2021ldu,Du:2021zzh,Albaladejo:2022sux,Mikhasenko:2022rrl}, it is possible that the observed states arise from compact bare states, that couple strongly to the continuum. In this respect, it is crucially important to know whether such compact states are expected or not from models at the quark level.

We find room for improvement  with respect to the use in \cite{Maiani:2019lpu} of the hyperfine $\kappa[(u d)_{\bf {\bar 3}}]$ coupling taken from baryon spectrum, the coupling which regulates the mass splitting of $\Sigma_Q$--$\Lambda_Q$ baryons. 
As demonstrated in previous cases,\footnote{See, Ref.~\cite{Maiani:2014aja} for the suppression in $Z_c(3900)$,~$Z_c'(4020)$ mass spectrum of $\kappa[(u \bar u)_{\bf{ 1}}]$ hyperfine coupling, dominant in meson spectra.} the extension to tetraquarks of hyperfine couplings taken from meson and baryon spectra is, in fact, an unjustified assumption. Hyperfine couplings depend crucially from the  overlap probability of the quark  pair involved, which, in tetraquarks cannot be {\it a priori} assumed to be equal to the overlap probabilities of the same pair  in mesons and baryons.

Recently, several studies of doubly heavy tetraquarks mass spectrum have been presented, based on the constituent quark model following the work of~\cite{Karliner:2017qjm,Luo:2017eub} (for an updated list, see~\cite{Guo:2021yws} and references therein). These analyses invariably use the hyperfine coupling taken from baryon spectrum, and fall under the same criticism.

We proceed to the calculation in the Born-Oppenheimer approximation in two ways: 

\begin{itemize}
\item {\bf{Method 1: scaling baryon and mesons hyperfine couplings with the dimensions of the BO bound state.} }We use the spin-independent BO formalism to evaluate the average separations of light quarks  and of heavy quarks. We obtain realistic estimates of the corresponding hyperfine couplings by scaling with respect to the separations in baryons (for $\bar q\bar q^\prime$) and in charmonium/bottomonium (for $QQ$).
\item {{\bf Method 2:  QCD approach.}} We include in the BO potential the contribution of the hyperfine QCD interaction at the quark level~\cite{DeRujula:1975qlm,Godfrey:1985xj,Capstick:1986ter}. Its first-order effect on the energy of the light quark system depends on the separation of the heavy sources, $R$, and it adds a contribution to the Born-Oppenheimer potential, which depends on the light quark spin $S_{\bar q\bar q}$ and on the total angular momentum $J$ of the tetraquark.\footnote{This method is followed in lattice calculations, where the computed Born-Oppenheimer potential takes full account of flavor and spin properties of the light quarks, see e.g.~\cite{Bicudo:2021qxj}.} The effect of the remaining heavy-to-heavy hyperfine interaction can be evaluated perturbatively, using the same formula  applied to the final wave function of the heavy quarks. 
\end{itemize}
This calculation leads to the following results:
\begin{enumerate} 
\item For the $I=0,~J^P=1^+$ state, the two methods give remarkably similar values, close to the observed mass of ${\cal T}_{cc}(3875)^+$. 
\item We compute  the masses  of the remaining, double charm states with $I=S_{\bar q\bar q}=1$ and $J^P=0^+,~1^+,~2^+$. Unlike the familiar $ \Lambda_Q,~ \Sigma_Q$ cases, the doubly heavy, $I=1,~J^P=1^+$ tetraquark is almost degenerate with the isoscalar partner. However, as discussed later, theory uncertainties allow for it to appear up to $20\mev$ below the $DD^*$ threshold, thus escaping detection at LHCb.
 \item Concerning the $[bb\bar q\bar q],~I=0$ tetraquark, the new evaluation gives a mass below the $ B B$ threshold  but rather close to it, not allowing a definite decision about the issue of stability against short-lived (strong and electromagnetic) decays.
 \end{enumerate}

{\bf \emph{Color couplings.}} In pursuing the analogy with the treatment of the  hydrogen molecule, the coulombic potential terms are rescaled by the appropriate color factors. Quarks are treated as non-relativistic  and weakly interacting. The determination of color factors is done in the one-gluon-exchange  approximation. 

As in~\cite{Maiani:2019lpu} we consider doubly flavored $bb$~and~$cc$ tetraquarks, with the doubly heavy pair in color $ \bm{\bar{3}}$.  
The lowest energy state corresponds to $QQ$ in spin one and light antiquarks in spin and isospin zero.  The tetraquark state is $|T\rangle=\left|(QQ)_{\bar {\bm 3}}, (\bar q\bar q)_{ {\bm 3}} \right\rangle_{\bm 1}$. From the Fierz identity
\be
|T\rangle=\sqrt{\frac{1}{3}}\left|(\bar q Q)_{\bm 1},(\bar q Q)_{\bm 1}\right\rangle _{\bm 1}-\sqrt{\frac{2}{3}}\left|(\bar q Q)_{\bm 8},(\bar q Q)_{\bm 8}\right\rangle _{\bm 1}\label{tetra3}
\ee
weighting with the squared amplitudes in~\eqref{tetra3}, one derives the attractive color factors \footnote{We use the rule based on quadratic Casimir coefficients $\lambda_{12}=1/2(C(\bm S)-C(\bm R_1)-C(\bm R_2))$ where $\bm S$ is one of the representations contained in the Kronecker product $\bm R_1\otimes \bm R_2$. $C(\bm 3)=C(\bar{\bm 3})=4/3$, $C(\bm 6)=10/3$ and $C(\bm 8)=3$. } 
\begin{align}
\lambda_{QQ}&=\lambda_{\bar q \bar q}=-\frac{2}{3}\alpha_s\notag\\
\lambda_{Q\bar q}&=\left[\frac{1}{3}\times\frac{1}{2}\left(-\frac{8}{3}\right)+\frac{2}{3}\times\frac{1}{2}\left(3-\frac{8}{3}\right)\right]\alpha_s=-\frac{1}{3}\alpha_s 
\label{bqbar3}
\end{align}
We shall add to the QCD coulombic potential a linearly rising, confining,  potential, $V=k_{Q\bar q}\, r$. The string tension $k_{Q\bar q}$ in the $Q\bar q$ orbital, is obtained from the charmonium string tension $k$ according to the so-called Casimir scaling~\cite{Bali:2000gf}
\be
k_{Q\bar q}=\frac{3}{4\alpha_s}|\lambda_{Q\bar q}|\, k =\frac{1}{4} k \label{kbqbar3}
\ee
where $k$ is the string tension derived from the charmonium spectrum where $|\lambda_{c\bar c}|=4/3 \alpha_s$.
 
As shown in~\eqref{tetra3}, $Q\bar q$ is in a superposition of color singlet and color octet. The charge of $(\bar q Q)_{\bm 8}$ is represented by an $SU(3)$ tensor $v^i_j$, traceless. In the QCD vacuum this charge might be neutralized by soft gluons, as in $A^j_iv^i_j$: in that case only the singlet component matters, and $k_{Q\bar q}=k$. We call this possibility `triality scaling'\footnote{Consider a generic color charge described by a  SU(3) tensor $v^{i_i\cdots i_n}_{j_1\cdots j_m}$, having triality ${\cal T}=n-m-3\lfloor (n-m)/3\rfloor$. It  can be lowered to $v^{i_1\cdots i_{n-m}}$ by repeated contraction with soft gluons $A^{j_m}_{i_n}$. If $n-m=1$ we get a $\bm  3$ tensor. If $n-m=2$ we get a $\bm 6$. If $n-m\geq 3$,  $v^{i_1\cdots i_{n-m}}$  can be further reduced by contraction with the $\overline{\bm{10}}$ tensors $A^r_{i_1}A^s_{i_2}\epsilon_{i_3 r s}$ ($i_1, i_2, i_3$ symmetrized) to finally get either one of  $\bm 1,\bm 3,\bm 6$.  Therefore the product of a charge $v^{i_i\cdots i_n}_{j_1\cdots j_m}$  and its conjugate can be reduced  to the non-trivial cases  $\bm 3\otimes \bar {\bm  3}$ as in~\eqref{tetra3},  or $\bm 6\otimes \bar{ \bm 6}$. The Kronecker decomposition of  $\bm 6\otimes \bm 8$ contains the $\bar{\bm 3}$ representation as well as $\bar{\bm 6}\otimes \bm 8$ contains the $\bm 3$. Therefore,  by the effect of the contraction with gluons, also $\bm 6\otimes \bar{\bm 6}$ behaves like $\bm 3\otimes \bar {\bm 3}$ and we still might use $k$ rather than the Casimir scaled value.}. We will show the results of both hypotheses for the string tension
\be
k_{Q\bar q}=\left\{\frac{k}{4},k\right\}
\label{kbqbar3-2}
\ee
indicated, respectively, as Casimir and triality scaling.

\begin{figure}[ht!]
   \centering
   \includegraphics[width=7truecm]{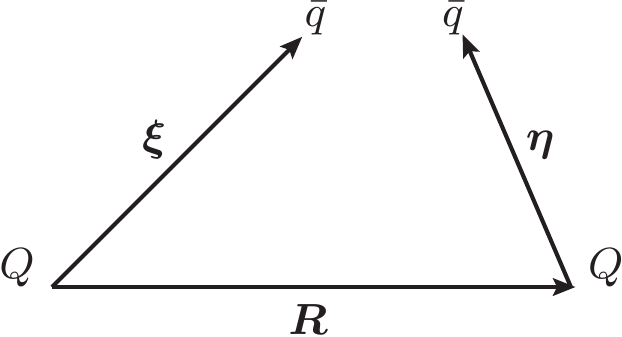}
   \caption{The heavy quarks are separated by the vector $\bm R$. The vectors $\bm \xi$ and $\bm \eta$ have their application points at the two heavy quarks. \label{fig1}}
\end{figure}

\vspace{.5cm}
{\bf \emph{Orbitals.}} We consider at first the heavy quarks as fixed color sources at a distance $R$.
 Light antiquarks are  bound each  to a heavy quark in orbitals with wave functions $\psi(\bm \xi)$ and $\phi(\bm \eta)$ and the ground state of the $\bar q \bar q$ system is assumed to be symmetric under the exchange of light quarks coordinates (the notation is defined in Fig.~\ref{fig1}).

\be
\Psi=\frac{\psi(\bm \xi)\phi(\bm \eta)+\psi(\bm \eta)\phi(\bm \xi )}{\sqrt{2\left[1+S^2(R)\right]}}\label{ground}
 \ee
 Normalization, $( \Psi, \Psi)=1$, is obtained with the overlap function given by\footnote{Considering ground states only, we restrict $\psi$ and $\phi$ to be real functions.}
\be
S(R)= \int_{\bm \xi} ~\psi(\bm \xi)\phi(\bm \xi)
\ee
The wave function $\psi(\bm \xi)$ gives the amplitude of $\bar q$ at a distance $\bm \xi$ from $Q$, as represented in Fig.~\ref{fig1}. 
The wavefunction $\phi(\bm \eta)$ is the amplitude of  the other light quark $\bar q$ at a distance $\bm \eta$ from the second heavy quark (which is at distance $\bm R$ from the former). The vectors $\bm \xi, \bm \eta$ have the application points in the positions of the  two heavy quarks respectively. The  $\psi$ and $\phi$ wavefunctions are written in terms of the radial functions ${\mathcal R}=R_{00}/\sqrt{4\pi}$  in the following way 
 \begin{align}
\psi(\bm \xi)&= {\cal R}(|\bm \xi|) & \psi(\bm \eta) &= {\cal R}(|\bm R+\bm \eta| )\notag \\
\phi(\bm \eta)&={\cal R}(|\bm \eta|) & \phi(\bm \xi)&={\cal R}(|\bm \xi -\bm R|)
 \label{cinque}
 \end{align}
${\cal R}(r)$ is the radial wave function obtained by solving variationally the Schr\"odinger equation of the heavy quark-light antiquark system with the potential,
\begin{align}
V(r)&= \frac{\lambda_{Q\bar q}}{r} + k_{Q\bar q} \,r + V_0 = -\frac{1}{3}\frac{\alpha_s}{r} +\frac{1}{4}k \, r  +V_0 \label{potorb}\\ 
{\cal R}(r)&=\frac{A^{3/2}}{\sqrt{\pi}}e^{-Ar} \label{sei}
\end{align}

We have included a constant $V_0$, to be discussed below, that defines the offset of the energy for confined systems. The determination of $A$ comes from the minimization of  $({\cal R},H\,{\cal R})=\langle H\rangle$: the value of $A$ used in computations corresponds to $\langle H\rangle_{\rm min}$. The light quarks energy, to zeroth order when we restrict to the interactions that define the orbitals, is
\be
{\cal E}_0=2(\langle H\rangle_{\rm min} +V_0 )
\label{eps0}
\ee
where  $\langle H\rangle_{\rm min}$ is the orbital energy eigenvalue (and the minimum of the Schr\"odinger functional).

In Ref.~\cite{Maiani:2019lpu} and in the following, we use the numerical values:
\begin{align}
\alpha_s(2M_c)&=0.30 
& \alpha_s(2M_b)&=0.21 &  k&=0.15\gevsq
 \label{bb&bc}
\end{align}

{\bf \emph{Determination of the BO potential.}}  We include  in a perturbation Hamiltonian the interactions left out from the construction of the orbitals, namely the interaction of each light quark with the other heavy quark and the interaction among light quarks. Following Fig.~\ref{fig1}
\be
\delta H=\lambda_{Q\bar q} \left(\frac{1}{|\bm \xi- \bm R|}+\frac{1}{|\bm \eta +\bm R|}\right) +\frac{\lambda_{q\bar q}}{|\bm \xi- \bm R-\bm \eta |}
\label{nove}
\ee
with color factors taken from~\eqref{bqbar3}.
We compute the total energy of the light system in the presence of fixed sources, ${\cal E}(R)$, to first order in $\delta H$
\begin{align}
{\cal E}(R)&={\cal E}_0 + \Delta E(R)\notag \\
\Delta E(R)&=(\Psi, \delta H \Psi)=\frac{1}{1+S^2(R)}\left[ -\frac{1}{3}\alpha_s ( 2I_1(R)+2S(R) I_2(R))-\frac{2}{3}\alpha_s(I_4(R) + I_6(R))\right]
\label{dieci}
\end{align}

The $I_i(R)$ are integrals over the orbital wave functions are defined and computed in~\cite{Maiani:2019lpu},\footnote{When computing e.g. $I_1$, the angle between $\bm \xi$ and $\bm R$ corresponds to the  polar angle $\theta$ in the $\bm \xi$ integration. The distance between light quarks  $|\bm \xi-\bm R-\bm \eta|=d_{\bar q\bar q}$, occuring in $I_{4,6}$ can be computed by shifting along $x$ or $y$ as in 
 $$d_{\bar q \bar q}= \sqrt{(\xi  \sin (\theta_\xi) \cos (\phi_\xi)-\eta  \sin (\theta_\eta ) \cos (\phi_\eta ))^2+(\xi  \cos (\theta_\xi)-\eta  \cos (\theta_\eta ))^2+(-\eta 
   \sin (\theta_\eta ) \sin (\phi_\eta )+\xi  \sin (\theta_\xi) \sin (\phi_\xi)-R)^2}$$ where the polar and azimuthal angles are related to $\bm \xi$ and $\bm \eta$. } 
\begin{align}
I_1(R)&\equiv\int_{\bm \xi }\psi(\bm \xi)^2\frac{1}{|\bm \xi-\bm R|}=\int_{\bm \eta}\phi(\bm \eta)^2\frac{1}{|\bm \eta+\bm R|}\notag \\
I_2(R)&\equiv\int_{\bm \xi }\psi(\bm \xi) \phi(\bm \xi) \frac{1}{|\bm \xi-\bm R|}=\int_{\bm \eta }\psi(\bm \eta) \phi(\bm \eta) \frac{1}{|\bm \eta+\bm R|}\notag \\
I_4(R)&\equiv\int_{\bm \xi,\bm \eta }\psi(\bm \xi)^2 \phi(\bm \eta)^2 \frac{1}{|\bm \xi-\bm R-\bm \eta|} = \int_{\bm \xi,\bm \eta }\psi(\bm \eta)^2 \phi(\bm \xi)^2 \frac{1}{|\bm \xi-\bm R-\bm \eta|}  \notag \\
I_6(R)&\equiv\int_{\bm \xi,\bm \eta }\psi(\bm \xi)\phi(\bm \xi) \psi(\bm \eta) \phi(\bm \eta) \frac{1}{|\bm \xi-\bm R-\bm \eta|}
\end{align}
Results in the first three lines are derived from the symmetry transformation $\bm\xi\to \bm\eta$, $\bm R\to -\bm R$, $\psi\to\phi$.  With these definitions at hand the result~\eqref{dieci} for $\Delta E(R)$ is readly derived from the definition~\eqref{nove} of $\delta H$.

The Born-Oppenheimer potential, to be used in the Scr\"odinger equation of the heavy quarks, is then
\be
V_{\rm BO}(R)=-\frac{2}{3}\alpha_s\frac{1}{R}+{\cal E}(R)\label{bopot}
\ee
At large separations $V_{\rm BO}(R)$ tends to the constant value 
 \be
 V_{\rm BO}(R)\to {\cal E}_0=2\left( \langle H\rangle_{\rm min} +V_0\right)\qquad \text{for } R\to \infty
 \ee 
 As noted in~\cite{Maiani:2019lpu}, at infinity the two orbitals tend to a superposition of color ${\bf 8}$--${\bf 8}$ and color  ${\bf 1}$--${\bf 1}$. The color of a triality zero pair can be screened by soft gluons from the vacuum, as first noticed in~\cite{Bali:2000gf} and supported by lattice QCD calculations (see~\cite{Bicudo:2021qxj} for recent results). The upshot is that, including the constituent quark rest masses taken from the meson spectrum, Tab.~\ref{mas1}, the limit $V_{\rm BO}(\infty)+2(M_Q + M_q)$ 
must coincide with the mass of a pair of non-interacting beauty (charmed) mesons with spin-spin interaction subtracted, which is just $2(M_Q + M_q)$. 
Thus, we derive the boundary condition
\be
\langle H\rangle_{\rm min} +V_0=0
\label{noconf}
\ee
which fixes $V_0$. 
\bigskip

\emph{\bf{Tetraquark spectrum and $Q$ values.}} 
The negative eigenvalue $E$ of the Schr\"odinger equation with $V_{\rm BO}(R)$ (including the condition on $V_0$ just found) is the binding energy associated with the BO potential. 
The masses of the lowest tetraquark with $[(QQ)_{S=1}(\bar q\bar q)_{S=0}]$ and of the pseudoscalar mesons $P=Q\bar q$ are 
\begin{align}
M(T)&=2(M_Q + M_q) + E+\frac{1}{2}\kappa_{QQ}-\frac{3}{2}\kappa_{\bar q \bar q}\\
M(P)&=M_Q + M_q -\frac{3}{2}\kappa_{Q\bar q}
\end{align}
The resulting $Q$-values with respect to the $PP$ thresholds are
\be
Q_{QQ}=M(T)-2M(P)=E+\frac{1}{2}\kappa_{QQ}-\frac{3}{2}\kappa_{\bar q \bar q}+3\kappa_{Q\bar q}\label{eqQkarl}
\ee
With the values in \eqref{bb&bc} and in Table~\ref{mas1} we obtained~\cite{Maiani:2019lpu}: 
\begin{align}
E&=-70~(-87)\mev\qquad\text{for}~cc, \notag \\
E&=-67~(-85)\mev\qquad\text{for}~bb,\label{eigenvalues}
\end{align}
where the first result assumes Casimir scaling, and the one in parenthesis assumes triality scaling. For the $I=0,~J^P=1$ state, the $Q$-values turned out to be~\cite{Maiani:2019lpu}
\begin{align}
Q_{cc}&= +7 \,(-10)\mev  \label{Qcc} \\
Q_{bb}&= -138 \,(-156)\mev  \label{Qbb}
\end{align}
To obtain \eqref{Qcc} and \eqref{Qbb} we used the hyperfine couplings obtained from meson and baryon spectra reported in Tabs.~\ref{mas1} and~\ref{spin}~\cite{Ali:2019roi}. As mentioned in the Introduction, this hypothesis needs a closer examination. 

\begin{table}[t]
\centering
    \begin{tabular}{|c|c|c|c|c|}
     \hline
Flavors& $q$ &  $s$ &  $c$ & $b$ \\ \hline
$M$(\nsmev) & $308$ & $484$ & $1667$ & $5005$ \\ \hline
\end{tabular}
 \caption{\footnotesize {Constituent quark masses from $S$-wave mesons~\cite{Ali:2019roi}, with $q=u,d$.}}
 \label{mas1}
\end{table}
\begin{table}[t]
\centering
    \begin{tabular}{|c|c|c|c|c|c|c|c|} 
    \hline
Mesons & $(q\bar q)_1$&$(q\bar s)_1$&  $(q\bar c)_1$&  $(s\bar c)_1$ & $(q \bar b)_1$& $(c\bar c)_1$ & $(b\bar b)_1$  \\ 
\hline
$\kappa$ (\nsmev) & $318$ & $200$  & $70$ & $72$ & $23$  & 56 &   30\\ 
\hline\hline
Baryons & $(qq)_{\bar 3}$ & $(q s)_{\bar 3}$  &  $(q c)_{\bar 3}$ &  $(s c)_{\bar 3}$& $(q b)_{\bar 3}$& $(c c)_3$ & $(b b)_3$   \\ 
\hline
$\kappa$ (\nsmev) & $98$ &$59$ &   $15$ & $50$ & $2.5$ & 28& 15   \\ 
\hline
 \hline
Ratio $\frac{\kappa_{MES}}{\kappa_{BAR}}$ & 3.2& 3.4  & 4.7 &1.6  & 9.2 & -- &-- \\
\hline
\end{tabular}
 \caption{\footnotesize {$S$-wave Mesons and Baryons: spin-spin interactions of the lightest quarks with the heavier flavours~\cite{Ali:2019roi}.  Values for $\kappa[(Q\bar Q)_1]$ are taken from the mass differences of ortho- and para-quarkonia. Following the one-gluon exchange prescription one then takes $\kappa[(QQ)_3]=1/2\kappa[(QQ)_1]$.
 }}
 \label{spin}
\end{table}

Within the  Born-Oppenheimer scheme we will improve this calculation with two methods, as mentioned in the Introduction.

{\bf \emph{Method~1: Hyperfine couplings by rescaling the overlap probabilities.}} 
The average distance of the light quarks as a function of $R$, the heavy quarks distance, is given by  the integral~\cite{Maiani:2019lpu}:
\be
d_{\bar q\bar q}(R)=\left(\Psi, |{\bm \xi}-\bm R-{\bm \eta}|\,  \Psi\right)= \int_{\bm \xi,\bm \eta }
\frac{\psi(\bm\xi)^2\phi(\bm\eta)^2+\psi(\bm\xi)\phi(\bm\xi)\psi(\bm\eta)\phi(\bm\eta)}{1+S^2(R)}\, \left|{\bm \xi}-\bm R-{\bm \eta}\right| \label{dist}
\ee
The average distance between light quarks in the tetraquark  is then given by
\be
\bar d_{\bar q\bar q}=\int dR ~\chi^2(R)\, d_{\bar q\bar q}(R)
\ee
where $ \chi(R)$ is the normalized radial wave function of the $QQ$ pair, solution of the Schr\"odinger equation in the Born-Oppenheimer potential $V_{BO}(R)$. In correspondence, we scale the hyperfine coupling in the tetraquark by rescaling $\kappa_{qq}$ in Tab.~\ref{spin} as with the inverse cube of $\bar d_{\bar q\bar q}$. 

The inverse radius of diquarks $[qq]$ in baryons is estimated in Ref.~\cite{Karliner:2017gml} from the electrostatic contributions to the isospin breaking mass differences of baryons. They quote a parameter $a$ from which the radius is derived according to
\be
a=\alpha \left\langle R_{[qq]}^{-1} \right\rangle \simeq 2.83\mev\Longrightarrow R_{[qq]}\simeq 2.58\gev^{-1}
\ee
This  leads to estimate the rescaled copuling 
\be
\kappa^\prime_{qq}=\kappa_{qq}\, \left(R_{[qq]}/\bar d_{\bar q\bar q}\right)^3
\ee
We proceed analogously for the hyperfine  $QQ$ coupling in the tetraquark, defining 
\be
\bar d_{QQ}=\int dR \, \chi^2(R)\, R
\ee
We find  a characteristic value of $\bar d_{cc}\approx 5\gev^{-1}$ and $\bar d_{bb}\approx 3\gev^{-1}$. Similar results for $\bar d_{\bar q\bar q}$. 
We scale with the quarkonium average radius $R_{Q\bar Q}$, obtained variationally from the wave function of the Cornell potential
\be
V(r)=-\frac{4}{3} \frac{\alpha_s(M_Q)}{r}+ k\, r
\ee 
to obtain
\be
\kappa^\prime_{QQ}=\kappa_{Q Q}~\left(R_{Q\bar Q}/\bar d_{QQ} \right)^3
\ee
with $\kappa_{QQ}$ from Tab.~\ref{spin}.

From the treatment of charmed baryons which can be found in~\cite{Karliner:2019lau} we extract
\be
R_{Qq} \simeq 2.64\gev^{-1}
\ee
A quark pair $Q\bar q$ in  $QQ\bar q\bar q$ has two alternatives: $A)$ $Q$ and $\bar q$ belong to the same orbital, and lie at an average distance $\bar d_{Q\bar q}^A$;  $B)$  $Q$ and $\bar q$  belong to different orbitals, being at a relative distance $\bar d_{Q\bar q}^B$ . One has to rescale the couplings by the appropriate distances, i.e.
\be
\kappa^\prime_{Q\bar q}=\frac{\kappa_{Q\bar q}}{4}~\left[\frac{1}{2} \left(R_{Qq}/{\bar d}_{Q\bar q}^A \right)^3+\frac{1}{2}\left(R_{Qq}/{\bar d}_{Q\bar q}^B \right)^3\right]
\ee
where $\kappa_{Q\bar q}$ is taken from Tab.~\ref{spin}, $1/4$ is the color factor of $Q\bar q$ in the tetraquark with respect to the meson, and  the average distances are
\begin{subequations}
\be
\bar d_{Q\bar q}^A(R)=\int dR ~\chi^2(R)\int_{\bm \xi}
\frac{\psi(\bm\xi)^2+\psi(\bm\xi)\phi(\bm\xi)}{1+S^2(R)}\, \left|{\bm \xi}\right|
\ee
\label{dist2}
and
\be
\bar d_{Q\bar q}^B(R)=\int dR ~\chi^2(R)\int_{\bm \xi}
\frac{\psi(\bm\xi)^2+\psi(\bm\xi)\phi(\bm\xi)}{1+S^2(R)}\, \left|{\bm \xi}-\bm R\right| 
\ee
\end{subequations}

The resulting $Q$-values with respect to the $PP$ thresholds are finally
\be
Q_{QQ}=E+\frac{1}{2}\kappa'_{QQ}+\kappa'_{\bar q \bar q}\left[S_{\bar q \bar q}(S_{\bar q \bar q}+1)-\frac{3}{2}\right]+\kappa'_{Q \bar q}\left[J(J+1) - S_{\bar q \bar q}(S_{\bar q \bar q}+1)- 2\right]+3\kappa_{Q\bar q}\label{eqQ}
\ee

{\bf \emph{Method~2: Hyperfine couplings from QCD.}} 
We start from the interaction Hamiltonian  at the quark level,
\be
H_{ij}=-\frac{ \lambda_{ij}}{M_iM_j}~\frac{8\pi}{3}~\bm S_i \bm \cdot \bm S_j~\delta^{3}(\bm x_i-\bm x_j)\equiv K_{ij}~\bm S_i \cdot \bm S_j~\delta^{3}(\bm x_i-\bm x_j)
\ee
 with $\lambda_{ij}$ given in Eq.~\eqref{tetra3}.
Following~\cite{Godfrey:1985xj}, the light quark interaction Hamiltonian is
\be
H_{\bar q\bar q}=K_{\bar q\bar q}~\bm S_{\bar q} \cdot\bm  S_{\bar q}~\delta^3(\bm x_1-\bm x_2)\label{hfli}
\ee
where $\bm x_1-\bm x_2$ is the distance between the light quarks. 
According to the $\delta^3$-function in~\eqref{hfli} we have that $\bm \eta =\bm \xi -\bm R$ and 
\be
\eta=\sqrt{\xi^2+R^2-2R\xi \cos\theta}
\ee
In particular we find
 \be
 V_{\bar q\bar q}(R)=( \Psi, H_{\bar q\bar q} \Psi) = \frac{8\pi \alpha_s}{9M_q^2}~\int_{\bm \xi} \frac{\psi({\bm \xi})^2~\phi({\bm R}-{\bm \xi})^2}{1+S^2(R)}\times~\left\{ \begin{array}{c}-3~(S_{\bar q\bar q}=0)\\ \\+1~(S_{\bar q\bar q}=1)\end{array}\right.
 \ee
In the heavy-light case we have  (with an obvious notation we distinguish the two heavy quarks as $A,B$ and the light quarks as $1,2$)
\be
H_{Q\bar q}=K_{Q\bar q}\Big[{\bm S}_A\cdot {\bm { S}}_1~\delta^3({\bm x}_A-{\bm x}_1)+{\bm S}_A\cdot {\bm  S}_2~\delta^3({\bm x}_A-{\bm x}_2) + (A\to B)\Big]=H_{A1} +H_{A2} + (A\to B)
\ee
Therefore
\be
( \Psi, H_{A1} \Psi)=
\frac{K_{Q\bar q}}{2\left[1+S^2(R)\right]}\cdot \Big[ \psi(0)^2+\psi(R)^2+2S~\psi(0)\psi(R)\Big](\bm S_A\cdot \bm S_1)
\ee
where we used the fact that $\bm \xi =0$ thus $\bm \eta=-\bm R$ (and $\phi(-\bm R)=\phi(\bm R)=\psi(\bm R)$ from~\eqref{cinque} and
\eqref{sei}). 

Adding all terms, one finds
\be
V_{Q\bar q}(R)=K_{Q\bar q}~\frac{ \psi(0)^2+\psi(R)^2+2S~\psi(0)\psi(R)}{2(1+S^2)}~ {\bm S}_{QQ} \cdot {\bm S}_{\bar q \bar q} 
\ee
We have
\be 
V_{Q\bar q}(R)=0\quad \text{for}\quad S_{\bar q\bar q}=0
\ee
whereas for $S_{\bar q\bar q}=1$ we have
\be
V_{Q\bar q}(R)=\frac{4\pi\alpha_s}{9M_q M_Q}~\frac{ \psi(0)^2+\psi(R)^2+2S~\psi(0)\psi(R)}{2\left[1+S^2(R)\right]}\times~\left\{\begin{array}{cc}-4~(J=0)\\-2~(J=1)\\+2~(J=2) \end{array}\right. 
\ee
Both $V_{\bar q\bar q}(R)$ and $V_{Q\bar q}(R)$ are added to $V_\text{BO}(R)$ in Eq.~\eqref{bopot} before solving the Schr\"odinger equation. Finally the contribution of the $QQ$ interaction is added perturbatively. The following equation replaces~\eqref{eqQkarl}
\be
Q_{QQ}=E+\frac{1}{2}\kappa''_{QQ}+3\kappa_{Q\bar q}\label{eqQisgur}
\ee
where $+3\kappa_{Q\bar q}$ comes from subtracting $2M_P$ as in~\eqref{eqQkarl} and 
\be
\kappa''_{QQ} = \frac{K_{QQ}}{2} \int\,\frac{1}{4\pi}\left(\frac{\chi(R)}{R}\right)^2 \delta^3\left({\bm R}\right) d^3 R=\frac{2\alpha_s}{9M_Q^2} \chi'(0)^2\label{kappaprimQQ}
\ee
We remark that, with this method, the energy value $E$ already incorporates the spin interactions of light-light and light-heavy quarks, that were added perturbatively in Method~1.

\begin{table}[b]
\centering
\begin{tabular}{||c|c|c|c|c|c|c||}
\hline
&  $\kappa'_{\bar q\bar q}$  &  $\kappa'_{QQ}$ & $\kappa'_{Q\bar q}$ & $E$ & $Q$-value & BO Mass \\ \hline
$cc$ & $+1.9~(+5.0)$ & $+0.4~(+0.7)$ & $+0.7~(+2.0)$ & $-70.3~(-86.8)$ & $+137.0~(+116.1)$ & $3872~(3851)$  \\ \hline
 $bb$ & $+2.7~(+8.6)$ & $+0.3~(+0.4)$ & $+3.0~(+1.1)$ & $-72.5~(-91.7)$ & $-7.4~(-35.5)$ & $10553~(10525)$  \\ \hline
\end{tabular}
\caption{\footnotesize Scaling of couplings, $S_{\bar q\bar q}=0$, $J=1$. All units are in\mev.  Numbers in parentheses correspond to the triality scaling. The $Q$-value is taken from the $PP$ meson pair threshold. The mass of the state is calculated by adding the $Q$-value to the physical mass of the $P$ meson pair.}
\label{tab:karliner-sqq0-J1-x}

\begin{tabular}{||c|c|c|c|c||}
\hline
   & $\kappa''_{QQ}$  & $E$ & $Q$-value & BO Mass \\ \hline
$cc$ &  $+1.2~(+2.0)$ &  $-74.8~(-100.2)$ & $+135.8~(+110.8)$ & $3871~(3846)$  \\ \hline
 $bb$ &  $+0.5~(+0.7)$ & $-77.3~(-107.4)$ & $-8.0~(-38.0)$ & $10552~(10522)$  \\ \hline
\end{tabular}
\caption{\footnotesize Couplings from QCD, $S_{\bar q\bar q}=0$, $J=1$. All units are in\mev. Note that $E$ includes the effect of light-heavy and light-light hyperfine interactions.
The contribution from $\kappa''_{Q Q}$ is to be added, as indicated in Eqs.~\ref{eqQisgur} and \ref{kappaprimQQ}.  
Numbers in parentheses correspond to the triality scaling.}
\label{tab:out-sqq0-J1-x}
\end{table}

\begin{table}[t]
\centering
\begin{tabular}{||c|c|c|c|c|c|c|c||}
\hline
& $J$  & $\kappa''_{QQ}$ & $E$ & $Q$-value & BO Mass \\ \hline
\multirow{3}{*}{$cc$} & $0$ & $+1.1~(+1.6)$ & $-77.3~(-113.9)$ & $+133.2~(+96.9)$ & $3868~(3832)$  \\ 
 & $1$ & $+1.1~(+1.5)$ & $-73.1~(-98.2)$ & $+137.5~(+112.5)$ & $3872~(3848)$  \\
 &$2$ & $+1.0~(+1.4)$ & $-64.6~(-67.1)$ & $+145.9~(+143.6)$ & $3881~(3879)$  \\ \hline
\multirow{3}{*}{$bb$}  & $0$ & $+0.5~(+0.6)$ & $-73.4~(-95.5)$ & $-4.2~(-26.2)$ & $10556~(10534)$  \\ 
 & $1$ & $+0.5~(+0.6)$ & $-72.2~(-91.1)$ & $-3.0~(-21.8)$ & $10557~(10538)$  \\ 
& $2$ & $+0.5~(+0.6)$ & $-69.5~(-82.5)$ & $-0.3~(-13.2)$ & $10560~(10547)$  \\ \hline
\end{tabular}
\caption{\footnotesize Couplings from QCD, $S_{\bar q\bar q}=1$. All units are in\mev. 
Numbers in parentheses correspond to the triality scaling. }
\label{tab:out-sqq1}
\end{table}

{\bf{\emph{Results for $\bm I=\bm S_{\bm \bar q\bar q}=0$.}}}  The comparison between Table~\ref{tab:karliner-sqq0-J1-x} (Method~1) and Table~\ref{tab:out-sqq0-J1-x} (Method~2) is encouraging. The difference between Casimir and triality scaling provide an estimate of the theory uncertainty $ \approx 20$--$25 \mev$.
There is a remarkable agreement between the two results on the  $Q$-values and the ${\cal T}_{cc}$ mass are well consistent with the mass value  ${\cal T}_{cc}^+(3875)$  observed by LHCb~\cite{LHCb:2021vvq,LHCb:2021auc}. For the ${\cal T}_{bb}$, we find
\be
M({\cal T}_{bb})= 10552~(10522)\mev
\label{prediz}
\ee
The $Q$-value of ${\cal T}_{bb}$ compares well to the recent lattice QCD determination
$Q=M({\cal T}_{bb})-2M(B)=-13^{+38}_{-30}\mev$~\cite{Bicudo:2021qxj}.

{\bf{\emph{$\bm I=\bm S_{\bm \bar q\bar q}=1$.}}} We report in Table~\ref{tab:out-sqq1} the results for isovector states, restricting to Method~2 for simplicity. We see that in the BO approximation all quarks are at higher average relative distances than in ordinary baryons.  
This translates in the fact that all hyperfine splittings are small. 

One may wonder why the $I=1,~J=1$ state, that is almost degenerate to  $I=0,~J=1$,  has not been seen by LHCb yet. The $T_{cc}(3875)^+$ is observed by LHCb over a large background of $pp$ collision  products. If the mass of the $I=J=1$ state is actually close to $T_{cc}(3875)^+$, its non observation  may be due to a significantly lower production cross section,  as it happens for the $\Sigma/\Lambda$ production ratio. If the mass falls outside the range $\approx -15$ to $\approx +5\mev$ from $T_{cc}$ (which is of the order of our theoretical uncertainty), the $I=J=1$ state would be out of the observational window of a $D D \pi$ line, even for a comparable cross section to the $I=0$ state. 

As for the other spin partners, neither could be seen in the $DD^*$ LHCb analysis ($J=0$ is forbidden, and $J=2$ decays in $D$-wave and is suppressed at threshold). However, both could be detected in $DD$.

Our results can be compared with other approaches in the literature, as the four-body calculation of~\cite{Richard:2018yrm,Hernandez:2019eox,Richard:2022fdc}, or with the global fits of~\cite{Luo:2017eub,Guo:2021yws,Karliner:2021wju}. The latter use invariably the hyperfine $\kappa[(u d)_{\bf {\bar 3}}]$ from the baryon spectrum, and predict large spin splittings, as well as a $\mathcal{T}_{bb}$ much lighter than our estimate. Detecting the bottom and spin partners of the $\mathcal{T}_{cc}$ will allow us to understand which method is capturing the right properties of multiquark systems.

{\bf \emph{Conclusions.}} 
We have presented the calculations  of   double-heavy tetraquarks,  based on a picture of the tetraquark system which is well described in  the Born-Oppenheimer approximation.  In this scheme the  mass of the ${\cal T}_{cc}$ state  is found with very good agreement with data. We predict a ${\cal T}_{bb}$ state, that agrees with lattice studies.  We find this result significant as the method used here is particularly simple.  

Our results are obtained following two different methods: $i)$ Scaling baryon and mesons hyperfine couplings with the dimensions of the Born-Oppenheimer  bound state: hyperfine couplings are scaled with respect to the separations in baryons, for $\bar q\bar q$, and in quarkonia, for $QQ$. Then the hyperfine couplings are included perturbatively  to the energy obtained solving the Schr\"{o}dinger equation with the Born-Oppenheimer potential $ii)$ Starting from the interaction Hamiltonian at the quark level, adding $V_{\bar q\bar q}(R)$ and $V_{Q \bar q}(R)$  to the Born-Oppenheimer potential and  then solving the Schr\"odinger equation.
The results obtained follwing these two distinct paths are in excellent agreement, adding solidity to the scheme used.

The average sizes of tetraquarks are estimated to be approximately $3$--$5\gev^{-1}$. At such distances the wave function is dominantly the meson-meson one, while at shorter distances it would be dominated by the diquark-antidiquark configuration, as illustrated by the Lattice QCD calculation in~\cite{Bicudo:2021qxj}. This can explain why the lineshape of such state gives a small (albeit negative) effective range. Finally, as a consequence of the larger size of tetraquarks with respect to ordinary baryons, hyperfine separations are not expected to be sizeable.   We discussed possible reasons why LHCb has observed  to date only one state in the $DD^*$ spectrum.

\acknowledgments

 We acknowledge interesting discussions with Ahmed Ali, Misha Mikhasenko, Giovanni Passaleva,  Marco Pappagallo and Alexis Pompili. We are indebted to Vanya Belyaev for valuable information about the current search of $I=J=1$ doubly charmed tetraquarks in LHCb. AP and ADP thank the CERN-TH Division for kind hospitality during the completion of this work.

\bibliography{quattro}

\end{document}